\documentclass{article}\usepackage{xspace}
\usepackage{tikz} 
\usepackage[utf8]{inputenc}
\usepackage{amsmath}
\usepackage{amssymb}
\usepackage{accenti}
 \usepackage{amsthm}
 \newtheorem{theorem}{Theorem}
  
 \newtheorem{lemma}[theorem]{Lemma}

\usepackage{url}
\usepackage{hyperref}
\usepackage{graphics} 
\usepackage{array}
\usepackage{multirow} 
\usepackage{float} 
\usepackage{subfigure}
\usepackage{faktor} 
\usetikzlibrary{snakes}
\usepackage{enumerate}
\usetikzlibrary{shapes,shapes.geometric,arrows,fit,calc,positioning,automata,decorations.pathmorphing}
\tikzset{state/.style={draw,ellipse}}

\DeclareMathOperator{\APOS}{POS}

\DeclareMathOperator{\First}{\mathsf{First}}

\DeclareMathOperator{\Last}{\mathsf{Last}}
\DeclareMathOperator{\Follow}{\mathsf{Follow}}

\DeclareMathOperator{\pos}{{\mathsf{Pos}}}

\DeclareMathOperator{\EdD}{\mathsf{Follow}}
\DeclareMathOperator{\EdS}{\mathsf{E}^\star}
\DeclareMathOperator{\Cross}{\mathsf{Cross}}

\DeclareMathOperator{\RE}{RE}
\DeclareMathOperator{\REna}{\mathsf{R}}
\def\first{\mathop{\textsf{f}}}
\def\last{\mathop{\textsf{s}}}
\def\cross{\mathop{\textsf{c}}}
\def\trans{\mathop{\textsf{t}}}
\def\edges{\mathop{\textsf{e}}}

\newcommand{\reg}{\alpha}
\newcommand{\regb}{\beta}
\newcommand{\regc}{\gamma}
\newcommand{\regp}{\alpha_P}
\newcommand{\nula}{\alpha_\varepsilon}
\newcommand{\nonnula}{\alpha_{\overline{\varepsilon}}}

\newcommand{\nulp}{\alpha_{P,\varepsilon}}
\newcommand{\nonnulp}{\alpha_{P,\overline{\varepsilon}}}

\newcommand{\letter}{\sigma}

\newcommand{\dfa}{DFA\xspace}
\newcommand{\dfas}{DFAs\xspace}
\newcommand{\nfa}{NFA\xspace}

\newcommand{\nfas}{NFAs\xspace}

\newcommand{\lang}{\mathcal{L}}

\newcommand{\apos}{\mathcal{A}_{\APOS}}

\newcommand{\eqre}{=}

\newcommand{\asympt}[1]{\;\lower6pt\hbox{$\stackrel{\mbox{\Large
        $\sim$}}{\mbox{\tiny{$#1$}}}$}\;}
\newcommand{\eG}{{\rm\Gamma}}
\newcommand{\lsbra}{[\![}
\newcommand{\rsbra}{]\!]}
\newcommand{\bras}[1]{\lsbra\,#1\,\rsbra}

\newcommand{\qee}{\makeatletter
\def\@eqnnum{{\normalfont \normalcolor \hbox{\qed}}}
\makeatother
}

\title{On the Uniform Distribution of \\Regular Expressions}
\author{
Sabine Broda \and 
António
  Machiavelo\and Nelma Moreira
\and Rog\'erio Reis \\
 CMUP \& DM-DCC, Faculdade de
  Ciências da Universidade do Porto,\\
  Rua do Campo Alegre, 4169-007 Porto, Portugal\\
\texttt{firstname.lastname@fc.up.pt}
}

\begin{document}
\maketitle
\begin{abstract} Although regular expressions do not correspond
  univocally to regular languages, it is still worthwhile to study
  their properties and algorithms. For the average case analysis one
  often relies on the uniform random generation using a specific
  grammar for regular expressions, that can represent regular
  languages with more or less redundancy. Generators that are uniform
  on the set of expressions are not necessarily uniform on the set of
  regular languages.  Nevertheless, it is not straightforward that
  asymptotic estimates obtained by considering the whole set of
  regular expressions are different from those obtained using a more
  refined set that avoids some large class of equivalent expressions.
  In this paper we study a set of expressions that avoid a given
  absorbing pattern. It is shown that, although this set is
  significantly smaller than the standard one, the asymptotic average
  estimates for the size of the Glushkov automaton for these
  expressions does not differ from the standard case.
\end{abstract}

\section{Introduction}

Average-case studies often rely on uniform random generation of
inputs.  In general, those inputs correspond to trees, and generators
are uniform on the set of these trees, but not on the set that those
inputs represent (such as languages or boolean functions). Koechlin et
al.~\cite{KoechlinNR19,KoechlinNR20} studied expressions that have
subexpressions which are (semantically) absorbing for a given
operator, calling them \emph{absorbing patterns}. For instance,
$(a+b)^\star$ is absorbing for the union of regular expressions over
the alphabet $\{a,b\}$, since $\reg+(a+b)^\star$, or
$(a+b)^\star+\reg$, is equivalent to $(a+b)^\star$ for any expression
$\reg$. After repeatedly applying the induced simplification, in the
example above replacing $\reg+(a+b)^\star$ by $(a+b)^\star$, the
resulting expression can be significantly smaller.
For uniformly random generated expressions of a given size, Koechlin
et al.~showed that the expression resulting from this simplification
has constant expected size.
That result led the authors to the conclusion that uniform random
generated regular expressions lack expressiveness, and in particular
that uniform distribution should not be used to study the average case
complexity in the context of regular languages.  This conclusion is
misleading in at least two aspects.
First, as pointed out above, one is considering regular expressions
and not regular languages themselves. For instance, if one wants to
estimate the size of automata obtained from regular expressions, one
disregards whether they represent the same language or not. What is
implied by the results of Koechlin et al.~is that, if one uniformly
random generates regular expressions, one cannot expect to obtain,
with a reasonable probability, regular languages outside a constant
set of languages. This means that a core set of regular languages have
so many regular expression representatives that the remaining
languages very scarcely appear.  While neither regular expressions
($\RE$) nor nondeterministic finite automata (\nfa) behave uniformly
when representing regular languages, it is known that deterministic
automata (\dfa) are a better choice, in the uniform model, as they are
asymptotically minimal~\cite{nicaud14}. In this sense, minimal \dfas{}
are a perfect model for regular languages. However, in practice,
regular expressions are usually preferred as a representation of
regular languages, and are used in a non-necessarily simplified form.
Moreover, all of these objects ($\RE$s, \nfas, and \dfas) are
combinatorial objects \textit{per se} that can have their behaviour,
as well as of the algorithms having them as input, studied on average
and asymptotically. One should not confuse regular expressions by
themselves with the languages that they represent.
Second, the results of Koechlin et al.~do not imply that asymptotic
estimates obtained by considering the whole set of regular expressions
are different from those obtained by using a more refined set with
less equivalent expressions.  For instance, some results obtained for
expressions in strong star normal form coincide with the ones for
standard regular
expressions~\cite{broda19:_averag_behav_regul_expres_stron}. In order
to further sustain the above claim, in this paper we consider the set
$\REna$ of regular expressions avoiding an absorbing pattern which
extends the pattern in the example above and was the one considered by
Koechlin et al. It is shown that, although the set $\REna$ is
significantly smaller than the set $\RE$, the asymptotic estimates for
the size of the Glushkov automaton on these sets is the same.
Given the complexity of the grammars expressing the classes here
studied, we had to deal with algebraic curves and polynomials of
degree depending on the size of the alphabet, $k$, which brought up
challenges that are new, as far as we know. Not only we had to use the
techniques developed in our previous work~\cite{Broda:2020aa}, but
also some non-trivial estimates using Stirling approximation, and some
asymptotic equivalence reductions in order to obtain the asymptotic
estimates, and their limits with $k$.

\section{The analytic tools} 
Given some measure of the objects of a combinatorial class,
$\mathcal{A}$, for each $n\in\mathbb{N}_0$, let $a_n$ be the sum of
the values of this measure for all objects of size $n$.  Now, let
$A(z)=\sum_n a_nz^n$ be the corresponding generating function
(\emph{cf}.~\cite{flajolet08:_analy_combin}).  We will use the
notation $[z^n]A(z)$ for $a_n$.  The generating function $A(z)$ can be
seen as a complex analytic function. When this function has a unique
dominant singularity $\rho$, the study of the behaviour of $A(z)$
around it gives us access to the asymptotic form of its coefficients.
In particular, if $A(z)$ is analytic in some indented disc
neighbourhood of $\rho$, then one has the following \cite[Corol.~VI.1,
p.~392]{flajolet08:_analy_combin}:
%
\begin{theorem}
  \label{thm:ApproxNewtonBinomial}
  The coefficients of the series expansion of the complex function
  $f(z) \asympt{z\!\to\! \rho} \lambda\,
  \left(1-\frac{z}{\rho}\right)^{\nu},$
  where $\nu\in\mathbb{C}\setminus\mathbb{N}_0$,
  $\lambda\in\mathbb{C}$, have the asymptotic approximation
  $[z^n]f(z) = \frac{\lambda}{\eG(-\nu)}\,n^{-\nu-1}\rho^{-n} +
  o\left(n^{-\nu-1} \rho^{-n}\right)$. Here $\eG$ is, as usual, the
Euler's gamma function and the notation
$f(z) \asympt{z\!\to\! z_0} g(z)$ means that
$\lim\limits_{z\to z_0} \frac{f(z)}{g(z)} = 1$.
\end{theorem}

\subsection{Regular Expressions}

Given an alphabet 
$\Sigma=\{\letter_1,\ldots,\letter_k\}$, the set~$\RE$ of (standard) \emph{regular expressions},
$\regb$, over~$\Sigma$ contains $\emptyset$ and the expressions
defined by the following grammar:
\begin{eqnarray}
  \regb &:=& \varepsilon \mid \letter \in \Sigma \mid (\regb +
               \regb) \mid (\regb \cdot \regb)  
               \mid (\regb^\star). \label{eq:rt}
\end{eqnarray}
The
\emph{language} associated to~$\regb$ is denoted by $\lang(\regb)$
and defined as usual (with $\varepsilon$ representing the empty word).  Two expressions $\regb_1$ and $\regb_2$ are
\emph{equivalent}, $\regb_1\eqre\regb_2$, if
$\lang(\regb_1)=\lang(\regb_2)$.  The \emph{(tree-)size} $|\regb|$ of
$\regb \in \RE$ is the number of symbols in $\regb$ (disregarding
parentheses).  The \emph{alphabetic size} $|\regb|_{\Sigma}$ is the number of letters occurring in $\regb$. 
The generating function of $\RE$ is $B_k(z)=\sum_{\regb\in \RE}z^{|\regb|}=\sum_{n>0} b_n z^n$, where $b_n$ is the number of expressions of size $n$,cf.~\cite{nicaud09:_averag_size_of_glush_autom,broda12:_averag_size_of_glush_and}. From grammar~\eqref{eq:rt} one gets $B_k(z)= (k+1)z+2zB_k(z)^2+zB_k(z)$.
Considering the quadratic equation this yields 
$
B_k(z)=\frac{1-z - \sqrt{1-2z-(7+8k)z^2}}{4z}.	
$ 
To use Theorem~\ref{thm:ApproxNewtonBinomial} one needs to obtain the singularity,
$\rho$, as well as the constants $\nu$ and $\lambda$. Following~Broda et al~\cite{broda12:_averag_size_of_glush_and,Broda:2020aa}, we have
\begin{equation*}
B_k(z) \asympt{z\!\to\!\rho_k} -
\frac{\sqrt{2-2\rho_k}}{4\rho_k} \left(1-\frac{z}{\rho_k}\right)^{\frac12},
\end{equation*}
where the singularity $\rho_k = \frac{1}{1+\sqrt{8+8k}}$ is the
positive root of $p_k(z)= 1-2z-(7+8k)z^2$.  Thus, applying
Theorem~\ref{thm:ApproxNewtonBinomial} and noting that
$\Gamma(-\frac{1}{2})=\sqrt{\pi}$, the number of expressions of size
$n$ is asymptotically given by
\begin{equation}
  \label{eq:znB(z)}
  [z^n] B_k(z) \asympt{n}
  \frac{\sqrt{2-2\rho_k}}{8\rho_k \sqrt{\pi}}\, 
  n^{-\frac32}\,\rho_k^{-n}, 
\end{equation}
where we use the notation $\asympt{n}$ instead of
$\asympt{n\!\to\!\infty}$.

\section{Regular Expressions without $\Sigma^\star$ in Unions}
We consider the set $\REna$ of all regular expressions $\reg$ such that
 $\Sigma^\star$ does not occur in an union.  Here $\Sigma^\star$ denotes any expression $(\letter_{i_1}+ \cdots +\letter_{i_k})^\star$ where $\letter_{i_1}, \ldots ,\letter_{i_k}$ is a permutation of $\Sigma$.
 Note that $\Sigma^\star$ represents an absorbing pattern in the sense of \cite{KoechlinNR19}, i.e. $(\reg + \Sigma^\star) = (\Sigma^\star + \reg) = \Sigma^\star$, and that $\REna$ still generates all regular languages over $\Sigma$. 
We first consider 
$\Sigma=\{a,b\}$, for which we have the following grammar $\mathcal G_2$ for $\REna$. 
\begin{eqnarray}
  \reg & := & \varepsilon \mid a \mid b \mid (\reg \cdot \reg) \mid (\reg^\star) \mid (\regp +\regp) \label{eq:reg} \\
  \regp& := & \varepsilon \mid a \mid b \mid (\reg \cdot \reg)  \mid
            (\reg_\Sigma^\star) \mid (\regp +\regp) \label{eq:regP}\notag \\ 
  \reg_\Sigma& := & \varepsilon \mid a \mid b \mid (\reg\cdot \reg) \mid (\reg^\star) \mid \regc \notag\\
  \regc & := & (\reg_{ab} + \reg_{ab}) \mid (\reg_{ab} +a) \mid (\reg_{ab} +b) \mid  (a+
           \reg_{ab}) \mid (b +\reg_{ab}) \mid (a+a) \mid (b+b)\notag \\ 
  \reg_{ab} &:=& \varepsilon \mid (\reg\cdot \reg) \mid (\reg_\Sigma^\star) \mid (\regp +\regp). \notag
\end{eqnarray}
The set of expressions generated by the nonterminals of $\mathcal G_2$,  are, respectively, 
\begin{eqnarray*}
\bras{\reg} &=& \REna,\\
	\bras{\regp} &=& \{\, \reg \in \REna \mid \reg \not= (a+b)^\star \wedge \reg \not= (b+a)^\star \,\},\\
	\bras{\reg_\Sigma} &=& \{\,\reg \in \REna \mid \reg \not= (a+b) \wedge \reg \not= (b+a) \,\},\\
	\bras{\regc} &=& \{\,(\reg_1+\reg_2) \in \REna \mid  \{\reg_1,\reg_2\} \not= \{a,b\}\,\},\\
	\bras{\reg_{ab}} &=& \{\,\reg \in \bras{\regp} \mid \reg \not= a \wedge \reg \not= b \,\}.
\end{eqnarray*}
In particular, we obtain the correctness of $\mathcal G_2$. 
\begin{lemma}\label{lem:grcorrect}
  An expression $\reg \in \RE$ is generated by $\mathcal G_2$ if and
  only the absorbing pattern $(a+b)^\star$ or $(b+a)^\star$ does not
  occur in a union.
\end{lemma}

Let $R_2(z)$ denote the generating function for the class $\REna$ when $|\Sigma|=2$. It follows from $\eqref{eq:reg}$ that 
$R_2(z) = 3z + zR_2(z)^2 + zR_2(z) + zR_P(z)^2,$
where $R_P(z)$ is the generating function for the class of expressions generated by $\regp$. Comparing $\bras{\reg}$ and $\bras{\regp}$, one observes that the only expressions not generated by $\regp$ are $(a+b)^\star$ and $(b+a)^\star$, which are both of size $4$. Thus, 
 $R_P(z) = R_2(z) -2z^4.$
In general, for an arbitrary alphabet $\Sigma = \{\letter_1,\ldots,\letter_k\}$, the expressions $\reg \in \REna$ satisfy the following grammar $\mathcal{G}_k$
\begin{eqnarray}
  \reg & := & \varepsilon \mid \letter_1\mid \cdots\mid \letter_k \mid (\reg \cdot \reg) \mid (\reg^\star) \mid (\regp +\regp), \label{eq:rena}
\end{eqnarray}
where 
$\bras{\regp}=\{\, \reg \in \REna \mid \reg \not= (\letter_{i_1}+ \cdots + \letter_{i_k})^\star \wedge \{\letter_{i_1}, \ldots ,\letter_{i_k} \}=\Sigma \,\}.$
As before, we obtain the following two equations for the corresponding generating functions, where $(k-1)! \binom{2k-2}{k-1}$ denotes the number of expression $(\letter_{i_1}+ \cdots + \letter_{i_k})^\star$ with  $\{\letter_{i_1}, \ldots ,\letter_{i_k} \}=\Sigma$, each of which has size $2k$.
\begin{eqnarray}
  R_k(z) &=& (k+1)z + zR_k(z)^2 + zR_k(z) + zR_{P,k}(z)^2,\label{eq:Rk0}\\
  R_{P,k}(z) &=& R_k(z) - (k-1)! \binom{2k-2}{k-1} z^{2k}.\label{eq:RP}
\end{eqnarray}
In the next section,  the asymptotic estimates of $[z^n]R_k(z)$ are computed.

\subsection{Asymptotic Estimates for the Number of Expressions in $\REna$}
\label{sec:r(z)}
The generating function $R_k= R_k(z)$ satisfies the following
equation:
\begin{equation}
  \label{eq:quadR}
  2 z R_k^2 - r_k R_k + z s_k = 0,
\end{equation}
where
\begin{eqnarray*}
  r_k= r_k(z) &=& 1-z+2z^{2k+1} C_k,\\
  s_k=s_k(z) &=& 1+k+z^{4k} C_k^2,\\
 C_k &=& \binom{2k-2}{k-1} (k-1)! = \frac{(2k-2)!}{(k-1)!}.
\end{eqnarray*}
The discriminant of equation~\eqref{eq:quadR} is
$\Delta_k=\Delta_k(z) = p_k(z) + 4 z^{2k+1} C_k h_k(z),$
where 

\begin{eqnarray*}
  p_k=p_k(z) &=& 1-2z-(7+8k)z^2,\\
  h_k=h_k(z) &=& 1-z-C_k\, z^{2k+1}.
\end{eqnarray*}
Thus,
\begin{equation}
  \label{eq:Rk}
  R_k=R_k(z) = \frac{r_k- \sqrt{\Delta_k}}{4z},
\end{equation}
where the choice of the sign is determined by noticing that
$r_k(0) = \Delta_k(0) = 1$.
Let us now show that $R_k(z)$ has a unique determinant singularity in
the interval $]0,1[$, for all $k$. The ideia is to use the fact that
the polynomial $p_k(z)$ has only one positive zero, namely
$\rho_k$, use Rouch\'e's Theorem to show
that, in the disk $|z|<\frac{1}{\sqrt{8+8k}}$, the polynomial
$\Delta_k(z)$ has exactly one root in that disk, and finally show that
that unique root is real.  We recall that Rouch\'e's Theorem states
that, in particular, for polynomials $f(z)$ and $g(z)$ such that
$|f(z)-g(z)| < |f(z)| + |g(z)|$ holds for all $|z| = R$, in the
complex plane, then $f(z)$ and $g(z)$ have the same number of roots,
taking into account multiplicities, in the disk
$|z|<R$~\cite[Thm~3.3.4]{simon15:_basic_compl_analy}.
In order to estimate $|\Delta_k(z)-p_k(z)|$, we start by noticing that
from Stirling approximation,
$\sqrt{2\pi}\, n^{n+\frac12} e^{-n} \leq n!\leq n^{n+\frac12}
e^{1-n}$, valid for all $n\in\mathbb{N}$, one gets that, for all
$k\geq 2$,
\begin{equation*}
  \frac{\sqrt{2\pi}\,
    (2k-2)^{2k-\frac32}\,e^{2-2k}}{(k-1)^{k-\frac12}\, e^{2-k}}
  \leq C_k = \frac{(2k-2)!}{(k-1)!}\leq
  \frac{(2k-2)^{2k-\frac32}\,e^{3-2k}}{\sqrt{2\pi}\,
    (k-1)^{k-\frac12}\, e^{1-k}},
\end{equation*}
i.e.,~
\begin{equation}
  \label{eq:EstCk}
  \frac{\sqrt{2\pi}\,2^{2k-\frac32}(k-1)^{k-1}}{e^k}
  \leq C_k \leq
  \frac{2^{2k-\frac32} (k-1)^{k-1}}{\sqrt{2\pi}\,e^{k-2}}.
\end{equation}
Therefore, for $|z|=\frac{1}{\sqrt{8+8k}}$, 
\begin{eqnarray*}
  |\Delta_k(z)-p_k(z)|
  &\leq& 4 C_k \frac{1}{(8+8k)^{k+\frac12}}\, |h_k(z)|\\
  &\leq& \frac{ (k-1)^{k-1}}{\sqrt{2\pi}\,e^{k-2}\,
         2^{k+1} (k+1)^{k+\frac12}}\left(1- \frac{1}{\sqrt{8+8k}} -
         \frac{C_k}{(8+8k)^{k+\frac12}}\right)\\
  &\leq& \frac{1.48}{(2e)^k(k-1)\sqrt{k+1}}\left(1- \frac{1}{\sqrt{8+8k}} -
         \frac{C_k}{(8+8k)^{k+\frac12}}\right).
\end{eqnarray*}
Noticing that, from~\eqref{eq:EstCk}, one has
$$\frac{\sqrt{2\pi}\, (k-1)^{k-1}}{e^k 2^{k+3} (k+1)^{k+\frac12}}\leq
\frac{C_k}{(8+8k)^{k+\frac12}} \leq
\frac{(k-1)^{k-1}}{\sqrt{2\pi}\,e^{k-2} 2^{k+3}(k+1)^{k+\frac12}}\,
,$$
one concludes that
$\lim\limits_{k\to\infty} |\Delta_k(z)-p_k(z)| = 0$.

Let us now find the minimum of $|p_k(z)|$ on the circunference
$|z|=\frac{1}{\sqrt{8+8k}} = R$. Put $z = R e^{i\theta}$. One has
\begin{eqnarray*}
  |p_k(z)|^2
  &=& |1-2R  e^{i\theta} - (7+8k) R^2 e^{2i\theta}|^2\\
  &=& (1-2R\cos\theta-(7+8k)R^2\cos
      2\theta)^2+(1-2R\sin\theta-(7+8k)R^2\sin 2\theta)^2\\
  &=& 2 + 4R^2 + (7+8k)^2 R^4 - 4R(\cos\theta+\sin\theta) - 2(7+8k)R^2
      (\cos 2\theta+\sin 2\theta)\\
  & &  + 4R^3 (7+8k) (\cos\theta \cos 2\theta+\sin\theta \sin
      2\theta)\\
  &=& 2 + \frac{1}{2+2k} + \left(\frac{7+8k}{8+8k}\right)^2 -
      \frac{2(\cos\theta+\sin\theta)}{\sqrt{2+2k}} - \frac{(7+8k)
      (\cos 2\theta+\sin 2\theta)}{4+4k}\\
  & & + \frac{(7+8k)(\cos\theta \cos 2\theta+\sin\theta \sin
      2\theta)}{4(k+1)\sqrt{2+2k}}.
\end{eqnarray*}
It follows that
$\lim\limits_{k\to\infty} |p_k(z)|^2 = 3 - 2
(\cos2\theta+\sin2\theta)$. Since
$\max\limits_{\theta} (\cos\theta+\sin\theta) = \sqrt{2}$, one
concludes that
$\lim\limits_{k\to\infty} |p_k(z)|^2 \geq 3 - 2\sqrt{2} > 0$.
From all this, one concludes that $|\Delta_k(z)-p_k(z)| < |p_k(z)|$
for large enough values of $k$, and so Rouch\'e's Theorem applies to
show that the polynomial $\Delta_k(z)$ has exactly one root in the
open disk $|z| < \frac{1}{\sqrt{8+8k}}$.
\footnote{It is actually true that $|\Delta_k(z)-p_k(z)|
< |p_k(z)|$ for all $|z| = \frac1{\sqrt{8+8k}}$ and $k\geq 2$.}
\medskip
Since $\Delta_k(0) = 1$, in order to show that that root must be real
it suffices to show that one has
$\Delta_k\left(\frac{1}{\sqrt{8+8k}}\right) <0$. This can be shown as
follows. Since
\begin{eqnarray*}
  \Delta_k\left( \frac1{\sqrt{8+8k}}\right)
  &=& 2^{-6 k-\frac{7}{2}}
      (k+1)^{-2 k-1} \left(2^{3 k+2} \left(4 \sqrt{k+1}-\sqrt{2}
      \right) (k+1)^kC_k\right.\\
  && \left.- 4 \sqrt{2} \, C_k^2- 64^k \left(8 \sqrt{k+1}-\sqrt{2}
     \right) (k+1)^{2 k}\right),
\end{eqnarray*}
we want to show that
$$
2^{3 k} \left(\sqrt{8k+8}-1 \right) (k+1)^kC_k < C_k^2+
2^{6k-2} \left(2\sqrt{8k+8}-1 \right) (k+1)^{2 k}.
$$
Using \eqref{eq:EstCk}, it is enough to show that
\begin{eqnarray*}
  \frac{2^k  e^{k+2} \left(\sqrt{8
  k+8}-1\right)}{\sqrt{\pi}}<
  2^{2k} \left(2 \sqrt{8
  k+8}-1\right)\frac{(k+1)^{k}}{(k-1)^{k-1}}e^{2k}+\pi
  \frac{(k-1)^{k-1}}{(k+1)^k}, 
\end{eqnarray*}
that follows from this trivially true inequality
\begin{eqnarray*}
  \frac{\sqrt{8k+8}-1}{\sqrt{\pi}}<
  2^k \left(2 \sqrt{8
  k+8}-1\right)\frac{(k+1)^{k}}{(k-1)^{k-1}}e^{k-2}.
\end{eqnarray*}

The singularity of $R_k(z)$ is therefore given by the unique root of
$ \Delta_k(z)$ in the interval $\left]0,\frac{1}{\sqrt{8k+8}}\right[$,
which will henceforth denote by $\eta_k$\label{eta}. It also follows from
Rouché's Theorem that this root has multiplicity one.  Now,
$\Delta_k(z) = \left(1-\frac{z}{\eta_k}\right) \psi_k(z)$, for some
$\psi_k(z) \in \mathbb{R}[z]$.  Using L'H\^opital's Rule, one has
\begin{equation}
  \label{eq:psik}
  \psi_k(\eta_k) = -\eta_k \Delta'_k(\eta_k).
\end{equation}
Then, one has 
$R_k(z) \asympt{z\!\to\!\eta_k}  \frac{-r_k(\eta_k) - \sqrt{\psi_k(\eta_k)}
  \left(1-\frac{z}{\eta_k}\right)^{\frac12}}{4\eta_k}.$
By Theorem~\ref{thm:ApproxNewtonBinomial}, one gets the following
asymptotic approximation for the number of regular expressions 
%
\begin{theorem}
  \label{thm:asymt_Rk}
  With the notations above, one has
  $$[z^n]R_k(z) \asympt{n}
  \frac{\sqrt{\psi_k(\eta_k)}}{8\eta_k\sqrt{\pi}} n^{-\frac32}
  \eta_k^{-n}.$$
\end{theorem}
Using~\eqref{eq:znB(z)}, we have
\begin{theorem}
  \label{thm:asymt_ratio}
  The asymptotic ratio of the number of expressions in $\REna$ and the
  number of expressions in $\RE$ is given by,
  $$\frac{[z^n] R_k(z)}{[z^n] B_k(z)} \asympt{n} \frac{
    \frac{\sqrt{\psi_k(\eta_k)}}{8\eta_k\sqrt{\pi}} n^{-\frac32}
    \eta_k^{-n}}{\frac{\sqrt{2-2\rho_k}}{8\rho_k \sqrt{\pi}}\,
    n^{-\frac32}\,\rho_k^{-n}} = \frac{\sqrt{\psi_k(\eta_k)}}
  {\sqrt{2-2\rho_k}} \left(\frac{\rho_k}{\eta_k}\right)^{n+1}.
  $$
\end{theorem}

Since, as seen before, $\eta_k > \rho_k$, for all $k$, this yields
that, for every $k$, this ratio tends to $0$ as $n\to\infty$. As such,
considering $\REna$ instead of $\RE$, actually avoids a significant
set of redundant expressions. Such an improvement, in the sense
of~\cite{KoechlinNR19}, might influence the results obtained by
asymptotic studies. In the following section we show that is not the case
 for the average  asymptotic size of the Glushkov automaton in
terms of states and
transitions~\cite{nicaud09:_averag_size_of_glush_autom,broda12:_averag_size_of_glush_and}.


\section{Asymptotic Average Size of the Glushkov Automaton}

The
Glushkov automaton~\cite{Glushkov:1961qr} is constructed from an equivalent  regular expression  $\regb$ using the set $\pos(\regb)$ of positions of the letters  in $\regb$, as the set of  states (plus one initial state). Let $\pos(\regb)=\{1,2,\ldots,|\regb|_\Sigma\}$,
$\pos_0(\regb)=\pos(\regb)\cup\{0\}$ and $\overline{\regb}$ denote
the expression obtained from $\regb$ by marking each letter with its
position in $\regb$.  The construction is based on the position sets 
$\First(\regb)=\{\, i\mid (\exists w) \ \letter_iw\in
\lang(\overline{\regb})\, \}$, $\Last(\regb)=\{\, i\mid
(\exists w) \ w\letter_i\in \lang(\overline{\regb})\, \}$, and
$\Follow(\regb)=\{\, (i,j)\mid (\exists u,v) \
u\letter_i\letter_jv\in \lang(\overline{\regb})\, \}$. The \emph{Glushkov automaton} for $\regb$ is
$\apos(\regb)=\langle \pos_0(\regb),\Sigma,\delta_{\APOS},0,F\rangle$ with
the set of transitions $\delta_{\APOS}=\{\, (0,\overline{\letter_j},j)\mid
j\in\First(\overline{\regb})\, \}\cup\{\, (i,\overline{\letter_j},j)\mid\;
(i,j)\in\Follow(\overline{\regb})\, \}$ and the set of final states
$F=\Last(\overline{\regb})\cup\{0\}$ if
$\varepsilon \in \lang(\regb)$, and $F=\Last(\overline{\regb})$,
otherwise.

In the next subsection,  we estimate the average number of letters in $\reg \in \REna$, i.e., the number of states of $\apos(\reg)$. In the last subsection we consider the number of transitions.
\subsection{Estimates for the Number of Letters}
\label{sec:letters}
The average number of letters  in  uniform random generated regular expressions of a given size have been estimated for different kinds of expressions~\cite{nicaud09:_averag_size_of_glush_autom,Broda:2020aa}. For standard regular expressions that value is half the  size of the expressions as the size of the alphabet goes to $\infty$. In the following we obtain the same value for expressions in $\REna$. To count the number of letters in all expressions of a given size we use the bivariate generating function
$\mathcal{L}_k(u,z) = \sum_{n,i\geq 1} c_{n,i} u^i z^n,$
where $c_{n,i}$ is the number of regular expressions of size $n$ with
$i$ letters. Therefore, the total
number of letters in all the regular expressions of size $n$ is given
by the coefficients of the sum of the two series
$$L_k (z) = \frac{\partial \mathcal{L}_k(u,z)}{\partial u}\bigg|_{u=1} =
\sum_{n,i\geq 1} i\, c_{n,i}\, z^n.$$

From grammar~\eqref{eq:rena} the  generating function $L_k(z)$ 
satisfies the following.
\begin{eqnarray}
  L_k(z) &=& kz + 2zL_k(z)R_k(z) +zL_k(z) +2zP_k(z)R_P(z), \label{eq:Lk}\\
  P_k(z) &=& L_k(z) - k! \binom{2k-2}{k-1} z^{2k}.\label{eq:Pk}
\end{eqnarray}

Using equations~\eqref{eq:Rk0},\eqref{eq:RP},\eqref{eq:Lk},\eqref{eq:Pk} and Buchberger's algorithm~\cite{Buchberger01} one obtains the following equation, which is satisfied by the generating function $L_k= L_k(z)$:
\begin{equation}
  \label{eq:quadL}
  \Delta_k L_k^2 + \bar{r}_k L_k - \bar{s}_k = 0,
\end{equation}
where
\begin{eqnarray*}
  \bar{r}_k &=& k z^{2k} C_k\, \Delta_k,\\
  \bar{s}_k &=&  k z^2+k^2 z^{2k+1}\, C_k\, 
                 \left((z-1) (1+2z^{4k+1} C_k^2)+2C_k(2+k)+2z^{6k+1}C_k^3      
                 \right).
\end{eqnarray*}
The discriminant of equation~\eqref{eq:quadL} can be shown to be
\begin{equation}
  \label{eq:Delta}
  \bar{\Delta}_k(z) = z^2 k^2 \Delta_k(z) g_k(z)^2,
\end{equation}
where 
\begin{equation}
  \label{eq:gk}
  g_k(z) = 2-C_k z^{2k-1} \left(h_k(z) - C_k
             z^{2k-1}\right).
\end{equation}
Therefore,
\begin{eqnarray*}
  L_k(z) &=& \frac{k z^{2k} C_k \Delta_k(z)\pm
             \sqrt{\bar{\Delta}_k(z)}}{2\Delta_k(z)} 
= \frac{k z^{2k} C_k}2 \pm 
             \frac{k z g_k(z)}{2\sqrt{\Delta_k(z)}}.
\end{eqnarray*}
Using the fact that we know $L_k'(0)=k$, one deduces that
\begin{equation}
  \label{eq:Lk}
  L_k(z) = \frac{k z^{2k} C_k}2 + 
  \frac{k z g_k(z)}{2\sqrt{\Delta_k(z)}}.
\end{equation}
Now, applying the procedure described in \textit{Broda et al.}~\cite{Broda:2020aa} one obtains:
\begin{theorem}
  \label{thm:asymt_Lk}
  With the same notations as above, where $\eta_k$ is as defined in page~\pageref{eta},
  $$[z^n]L_k(z) \asympt{n} \frac{k\, \eta_k\,
    g_k(\eta_k)}{2\sqrt{\pi}\sqrt{\psi_k(\eta_k)}}n^{-\frac12}
  \eta_k^{-n}.$$
\end{theorem}

Therefore, from Theorems~\ref{thm:asymt_Rk} and \ref{thm:asymt_Lk},
one deduces:
\begin{theorem}
  The asymptotic ratio of letters in the expressions in $\REna$
 is given by
  $$ 
  \frac{[z^n]L_k(z)}{n[z^n]R_k(z)} \asympt{n} \frac{4 k\,
    \eta_k^2\, g_k(\eta_k)}{\psi_k(\eta_k)}.
  $$
\end{theorem}
Let us now see that
\begin{equation}
  \label{eq:lim_Eta_k}
  \lim_{k\to\infty} k\,\eta^2_k  = \frac18.
\end{equation}
Since we know that $\Delta_k(0)=1$, and $\Delta_k(x)$ has exactly one
real root in the interval $\left[0,\frac1{\sqrt{8+8k}}\right]$, in
order to show that $\eta_k > \rho_k$ for all $k$, it is enough to show
that:
$$\Delta_k (\rho_k) = p_k(\rho_k) + 4 \rho_k^{2k+1} C_k h_k(\rho_k) >0,
\text{ i.e. } h_k(\rho_k)>0.$$
Now, $h_k(\rho_k)>0 \iff 1 > \rho_k + C_k \rho_k^{2k+1}\iff
\sqrt{8+8k} > \frac{C_k}{(1+\sqrt{8+8k})^{2k}}.$
From \eqref{eq:EstCk} it follows that
\begin{equation*}
  \frac{C_k}{(1+\sqrt{8+8k})^{2k}} \leq
  \frac{2^{2k-\frac32} (k-1)^{k-1}}{\sqrt{2\pi}\,e^{k-2}
    (1+\sqrt{8+8k})^{2k}}.
\end{equation*}
It is therefore enough to show:
\begin{equation*}
  \frac{2^{2k-\frac32} (k-1)^{k-1}}{\sqrt{2\pi}\,e^{k-2}
    (1+\sqrt{8+8k})^{2k}} < \sqrt{8+8k},
\end{equation*}
which is equivalent to
\begin{equation*}
  2^{2k-\frac32} (k-1)^{k-1} < \sqrt{2\pi}\,e^{k-2}
  (1+\sqrt{8+8k})^{2k} \sqrt{8+8k}.
\end{equation*}
This is the same as
\begin{equation*}
  \left(\frac{4}{e}\right)^{k} (k-1)^{k-1} <
  \frac{2^{\frac32}\,\sqrt{2\pi}}{e^2}\,
  (1+\sqrt{8+8k})^{2k} \sqrt{8+8k},
\end{equation*}
which follows from:
\begin{equation*}
  \left(\frac{4}{e}\right)^{k} (k-1)^k <
  \frac{2^{\frac32}\,\sqrt{2\pi}}{e^2}\,
  2^{2k+1} (2+2k)^{k+\frac12}.
\end{equation*}
That is obvious when rewritten as
\begin{equation*}
  \left(\frac{4}{e}\right)^{k} (k-1)^k <
  \left(\frac{2^{\frac32}\,\sqrt{2\pi}}{e^2}\,
  2\right)\, 4^k (2+2k)^{k+\frac12}.
\end{equation*}
Thus, we conclude that
\begin{equation}
  \label{eq:EtaBounds}
  \rho_k = \frac1{1+\sqrt{8+8k}} < \eta_k < \frac1{\sqrt{8+8k}}. 
\end{equation}
From this it immediately follows that
$\lim\limits_{k\to\infty} k\,\eta^2_k = \frac18$, and then
$\lim\limits_{k\to\infty} p_k(\eta_k)=0$. Using the right hand inequality in \eqref{eq:EstCk} together with
\eqref{eq:EtaBounds}, it is not hard to show the following result.
\begin{lemma}\label{lemma:asymptCk}
  For all $t,s\in\mathbb{R}$, one has
  \begin{equation}
    \lim_{k\to\infty} C_k k^t \eta_k^{2k+s} = 0. 
  \end{equation}
\end{lemma}
From all this, and from \eqref{eq:gk} and \eqref{eq:psik}, one easily
gets
$\lim\limits_{k\to\infty} g_k(\eta_k) = \lim\limits_{k\to\infty}
\psi_k(\eta_k) = 2$, and thus:
\begin{equation}
  \label{eq:UmEmeio}
  \lim_{k\to\infty} \frac{4 k\,
    \eta_k^2\, g_k(\eta_k)}{\psi_k(\eta_k)} = \frac12.
\end{equation}
This means that the following result holds.
\begin{theorem}\label{thm:UmEmeio}
  In regular expressions without
  $\Sigma^\star$ in unions, the asymptotic ratio of letters goes to $\frac12$ as $k$ goes to $\infty$.
\end{theorem}

\subsection{Estimates for the Number of Transitions}
\label{sec:trans}

The transitions of the Glushkov automaton are defined using the sets of positions
$\First$, $\Last$ and $\Follow$.  These sets can be inductively define
for $\reg \in \REna$, as it is usually
done~\cite{broda12:_averag_size_of_glush_and}. Let $\nula\in \REna$ be
the set of expressions such that $\varepsilon\in\lang(\nula)$ and let
$\nonnula$ represent the set of expressions such that
$\varepsilon\notin\lang(\nonnula)$. 
We have that those sets satisfy the following grammars:
\begin{eqnarray}
  \nula & := & \varepsilon \mid (\nula \cdot \nula) \mid (\reg^\star) \mid (\nulp + \regp ) \mid (\nonnulp + \nulp )\label{eq:nula}\\
\nonnula &:= & \letter\in \Sigma \mid (\nonnula\cdot \reg) \mid (\nula\cdot\nonnula ) \mid (\nonnulp +\nonnulp ),
\end{eqnarray}
\noindent where $\nulp$ and $\nonnulp$ represent the expressions $\regp$ such that $\varepsilon\in\lang(\nulp)$ and $\varepsilon\notin\lang(\nonnulp)$, respectively. Note that then $\reg \in \in \REna$ could be defined by: $\reg := \nula \mid \nonnula$.
With this we have the following definitions.
\begin{eqnarray*}
\begin{array}{lcl}
   \First(\varepsilon)& = &\emptyset,\\
  \First(\sigma_i) & = & \{i\},\\
  \First(\reg^\star) & = & \First(\reg),
\end{array}&\ \ &
\begin{array}{lll}
  \First(\regp + \regp') & = & \First(\regp) \cup \First(\regp'), \\
  \First(\nula\cdot\reg) & = & \First(\nula) \cup \First(\reg),\\
  \First(\nonnula\cdot\reg)&=&\First(\nonnula). 
\end{array}
\end{eqnarray*}
The definition of $\Last$ is almost identical and differs only for the
case of concatenation, which is
$ \Last(\reg\cdot\nula) = \Last(\reg) \cup \Last(\nula)$ and
$ \Last(\reg\cdot\nonnula) = \Last(\nonnula)$. Following Broda et
al. \cite{broda12:_averag_size_of_glush_and} the set $\Follow$
satisfies
\begin{equation*}
  \label{eq:EdD}
  \begin{aligned}
 \EdD(\varepsilon)&= \EdD(\sigma_i) = \emptyset, \\
    \EdD(\regp + \regp') &= \EdD(\regp) \cup \EdD(\regp'), \\
    \EdD(\reg\cdot\reg') &= \EdD(\reg) \cup \EdD(\reg') \cup
    \Last(\reg) \times \First(\reg'), \\
    \EdD(\reg^\star) &= \EdS(\reg), \  \text{where} \\
 \EdS(\varepsilon)& = \emptyset, \ \     \EdS(\sigma_i) =  \{(i,i)\}, \ \  \EdS(\reg^\star) =  \EdS(\reg),
\\
    \EdS(\regp + \regp') &= \EdS(\regp) \cup
    \EdS(\regp') \cup \Cross(\regp,\regp'),\\
    \EdS(\nula\cdot\nula') &=
      \EdS(\nula)\;\cup\; \EdS(\nula')\; \cup\;
      \Cross(\nula,\nula'),\\
      \EdS(\nula\cdot\nonnula') &=    \EdD(\nula)\; \cup\;\EdD^\star(\nonnula')\;\cup \; \Cross(\nula,\nonnula'),\\
 \EdS(\nonnula\cdot\nula') &=    \EdD^\star(\nonnula)\; \cup\;\EdD(\nula')\;\cup \; \Cross(\nonnula,\nula'),\\
 \EdS(\nonnula\cdot\nonnula') &=    \EdD(\nonnula)\; \cup\;\EdD(\nonnula')\;\cup \; \Cross(\nonnula,\nonnula'),\\
          \end{aligned}
  \end{equation*}
  \noindent with $\Cross(\reg,\reg') = \Last(\reg)
  \times \First(\reg')\cup 
  \Last(\reg')\times \First(\reg)$. 

The generating functions for $\nula$ and $\nonnula$, respectively, $R_{\varepsilon,k}(z)=R_{\varepsilon,k}$ and $R_{\overline\varepsilon,k}(z)=R_{\overline\varepsilon,k}$, satisfy
\begin{eqnarray*}
	R_{\varepsilon,k}& = &z +zR_{\varepsilon,k}^2+zR_k+2zR_{P,\varepsilon,k}R_{P,k}-zR_{P,\varepsilon,k}^2,\\
	R_{\overline\varepsilon,k}&= &R_{k}- R_{\varepsilon,k},\\
	R_{P,\overline{\varepsilon},k} &=& R_{\overline\varepsilon,k},\\
	R_{P,\varepsilon,k}&=& R_{P,k}- R_{\overline\varepsilon,k} = R_{\varepsilon,k} - C_k z^{2k}.
\end{eqnarray*}

From that we conclude that
\begin{eqnarray}\label{eqn:reps}
	R_{\varepsilon,k}&=& z+zR_k+zR_{P,k}^2-zR_k^2+2zR_{\varepsilon,k}R_k
\end{eqnarray}

The function that counts the cardinality of
${\First}(\reg)$ is  $\first(\reg)$ and is defined as follows:
\begin{equation}
\label{edDcounting}
\begin{aligned}
  \first(\letter_i) & = 1 ,\\
  \first(\regp + \regp') & =  \first(\regp) + \first(\regp'),\\
  \first(\nula\cdot\reg') & = \first(\nula) + \first(\reg'),\\
	\first(\nonnula\cdot\reg') & = \first(\nonnula),\\
  \first(\reg^\star) & =\first(\reg),
\end{aligned}
\end{equation}
Note that $\first((\letter_{i_1}+ \cdots + \letter_{i_k})^\star)=k$ for any permutation $\letter_{i_1}, \ldots, \letter_{i_k}$ of $\Sigma=\{\letter_1,\ldots,\letter_k\}$.  The correspondent generating function $F_k(z)=\sum_{\reg}\first(\reg)z^{|\reg|}=F_k$  satisfies the following equations
\begin{eqnarray*}
  \label{eq:LF}
 F_k& =& kz+zF_k+2zF_{P,k}R_{P,k}+zF_kR_{\varepsilon,k}+zF_kR_k,\\
F_{P,k}&=&F_k-kC_kz^{2k},\\
R_{\varepsilon,k}& =& z+zR_k+2zR_{\varepsilon,k}R_k+zC_k^2z^{4k}-2zR_kC_kz^{2k},
\end{eqnarray*}
 Let $\last(\reg)$ be the function that counts  the cardinality of
 ${\Last}(\reg)$ and $S_k(z)$ the correspondent generating function. By symmetry we have that $S_k(z)=F_k(z)$. The functions counting the cardinalities of  $\EdD(\reg)$  and $\EdS(\reg)$ are $\edges(\reg)$ and $\edges^\star(\reg)$, respectively. Those functions are defined as follows:\begin{equation}
\label{edDcounting}
\begin{aligned}
  \edges(\sigma) & =  \edges(\varepsilon) =  0,\\
  \edges(\regp + \regp') & =  \edges(\regp) + \edges(\regp'),\\
  \edges(\reg\cdot\reg') & = \edges(\reg) + \edges(\reg') + 
  \last(\reg)\first(\reg'),\\
  \edges(\reg^\star) & = {\edges}^\star(\reg),
\end{aligned}
\end{equation}
where ${\edges}^\star(\reg)$ is given by
\begin{equation*}
  \begin{aligned}
   {\edges}^\star(\varepsilon) & = 0, \text{\hphantom{mm}}
    {\edges}^\star(\sigma)  =  1,\\
    {\edges}^\star(\regp + \regp') & = {\edges}^\star(\regp) +
    {\edges}^\star(\regp') + \cross(\regp,\regp'),\\
    {\edges}^\star(\reg_\varepsilon\cdot\reg'_\varepsilon) & =
    {\edges}^\star(\reg_\varepsilon) +
    {\edges}^\star(\reg'_\varepsilon) +
    \cross(\reg_\varepsilon,\reg'_\varepsilon), \\
    {\edges}^\star(\reg_{\overline{\varepsilon}}\cdot\reg'_\varepsilon)
    & = {\edges}^\star(\reg_{\overline{\varepsilon}}) +
    \edges(\reg'_\varepsilon) +
    \cross(\reg_{\overline{\varepsilon}},
    \reg'_\varepsilon), \\
    {\edges}^\star(\reg_{{\varepsilon}}\cdot
    \reg'_{\overline{\varepsilon}}) & = \edges(\reg_{{\varepsilon}})
    + {\edges}^\star(\reg'_{\overline{\varepsilon}}) +
    \cross(\reg_{{\varepsilon}},\reg'_{\overline{\varepsilon}}), \\
    {\edges}^\star(\reg_{\overline{\varepsilon}}\cdot
    \reg'_{\overline{\varepsilon}}) & =
    \edges(\reg_{\overline{\varepsilon}}) +
    \edges(\reg'_{\overline{\varepsilon}}) +
    \cross(\reg_{\overline{\varepsilon}},
    \reg'_{\overline{\varepsilon}}),
  \end{aligned}
  \label{eq:EStar}
\end{equation*}

 with $\cross(\reg,\reg') = \last(\reg) \first(\reg') +
\last(\reg') \first(\reg)$.  
From the above the corresponding generating functions $E_k(z)=\sum_{\reg}\edges(\reg)z^{|\reg|}=E_k$ and $E^\star_k(z)=\sum_{\reg}\edges^\star(\reg)z^{|\reg|}=E_k^\star$, respectively, satisfy the following equations.
%
\begin{eqnarray*}
	E_k&=&2zE_{P,k}R_{P,k}+2zE_kR_k+zF_k^2+zE^\star_k,\\
    E^\star_k&=&kz + 
2zE_{P,k}^\star R_{P,k}+2zF_{P,k}^2 +
   2zE^\star_{\varepsilon,k}R_{\varepsilon,k}+2zF_{\varepsilon,k}F_{\varepsilon,k}\\
& & +\ zE^\star_{\overline\varepsilon,k}R_{\varepsilon,k} + zE_{\varepsilon,k}R_{\overline\varepsilon,k} + 2zF_{\varepsilon,k}F_{\overline\varepsilon,k} +
 zE_{\varepsilon,k}R_{\overline\varepsilon,k} \\
 & & + \ zE^\star_{\overline\varepsilon,k}R_{\varepsilon,k} + 2zF_{\varepsilon,k}  F_{\overline\varepsilon,k} 
+  2z E_{\overline\varepsilon,k}R_{\overline\varepsilon,k}  +2zF_{\overline\varepsilon,k}F_{\overline\varepsilon,k} +
zE^\star_k
\\ 
&=& kz+2zE_{P,k}^\star R_{P,k}+2zE^\star_kR_{\varepsilon,k}
+ 2zE_k(R_{k}-R_{\varepsilon,k})
\\& & +\ 2zF_{P,k}^2+ 2zF_{k}^2+zE^\star_k,\\
E_{P,k}&=& E_k- k^2C_kz^{2k},\\
E^\star_{P,k}&=& E^\star_k- k^2C_kz^{2k}.
\end{eqnarray*}

The last two equations follow from the fact that
$\edges((\letter_{i_1}+ \cdots +
\letter_{i_k})^\star)=\edges^\star((\letter_{i_1}+ \cdots +
\letter_{i_k})^\star)=k^2$, for any permutation
$\letter_{i_1}, \ldots, \letter_{i_k}$ of $\Sigma$. The cost function
$\trans(\reg)= \first(\reg) + \edges(\reg)$ computes the number of
transitions in the Glushkov automaton of $\reg$.  The generating
function associated to $\trans$ is given by
$T_k(z) = F_k(z) + E_k(z)$.
%
%
Setting $w=T_k(z)$, one has
$$c_2 w^2 + c_1 w + c_0 =0,$$
where the $c_i = c_i(k,z)$. Therefore,
$$w = \frac{-c_1 \pm \sqrt{c_ 1^2-4c_0 c_2}}{2 c_2}.$$
Now, one can see that
$c_1 = \Delta_k s_k$, $c_2 = \Delta_k a_k b_k^2$ and $c_1^2-4c_0 c_2 = k^2 \Delta_k q_k^2,$
from which it follows that
$$w = -\frac{s_k}{2a_k b_k^2} \pm \frac{k q_k}{2a_k b_k^2
  \sqrt{\Delta_k}}.$$

With $\eta_k$ as defined in p.\pageref{eta}, one can now deduce, as
above, that
$$T_k(z) \asympt{z\to\eta_k} \frac{k q_k(\eta_k)}{2 a_k(\eta_k) b_k(\eta_k)^2
  \sqrt{\psi_k(\eta_k)}}\left(1-\frac{z}{\eta_k}\right)^{\frac12},$$
and therefore
$$[z^n]T_k(z) \asympt{n} \frac{k q_k(\eta_k)}{2\sqrt{\pi} a_k(\eta_k)
  b_k(\eta_k)^2 \sqrt{\psi_k(\eta_k)}}\,\eta_k^{-n} n^{-\frac12}.$$

From all this, one gets:
$$\frac{[z^n]T_k(z)}{[z^n]R_k(z)} \asympt{n} \frac{4 k\eta_k
  q_k(\eta_k)}{ a_k(\eta_k) b_k(\eta_k)^2 \psi_k(\eta_k)}\, n.$$

With the help of a symbolic and numeric computing system one can
explicitly find out the polynomials\footnote{These polynomials are
  quite large, \textit{e.g.}~$q_k$ has $437$ monomials and degree
  $10+28k$.}  $a_k$, $b_k$, $q_k$, and then reducing them modulo
$\Delta_k$ (which has $\eta_k$ as a root), and then using
Lemma~\ref{lemma:asymptCk} and \eqref{eq:lim_Eta_k}, one obtains:
$$ a_k(\eta_k) \asympt{k} \frac12 k\eta_k\quad;\quad 
b_k(\eta_k)\asympt{k} \frac18 k\eta_k\quad;\quad q_k(\eta_k)
\asympt{k} \frac1{2048} k.$$
This yields
$$\lim_{k\to\infty} \frac{4 k\eta_k
  q_k(\eta_k)}{ a_k(\eta_k) b_k(\eta_k)^2 \psi_k(\eta_k)} = 1.$$

We have thus obtained the following result.
\begin{theorem}\label{thm:Um}
  For expressions of size $n$ over an alphabet of size $k$, the number
  of transitions in the Glushkov automaton for regular expressions,
  without $\Sigma^\star$ in unions, is asymptotically, with respect to
  $n$, given by $\lambda_k n$, where
  $\lim\limits_{k\to\infty} \lambda_k = 1$.
\end{theorem}

To grasp the progression of $\lambda_k$, observe that
$\lambda_2=4.03$, $\lambda_5=2.91$, $\lambda_{10}=2.30$,
$\lambda_{10}=1.89$, $\lambda_{50}=1.54$, $\lambda_{100}=1.38$,
$\lambda_{10000}=1.03$.
Theorems \ref{thm:UmEmeio} and \ref{thm:Um} show that the size of the
Glushkov automaton, both in states and transitions, is, on average and
asymptotically, independent of whether we consider all regular
expressions or the restricted set $\REna$ mentioned by Koechlin et al.

\paragraph{Acknowledgments}
This work was
    partially  supported by CMUP, 
  through FCT – Fundação para a Ciência e a Tecnologia, I.P., under the
  project with reference UIDB/00144/2020.

\bibliographystyle{splncs04}

%
\end{document}